\documentclass[preprint,showpacs,aps]{revtex4}

\newcommand {\bea}{\begin{eqnarray}}
\newcommand {\eea}{\end{eqnarray}}
\newcommand {\be}{\begin{equation}}
\newcommand {\ee}{\end{equation}}
\newcommand {\pslash}{p\!\!\!/}
\usepackage{graphicx}
\usepackage{epsfig,subfigure}
\usepackage{amsmath}
\begin{document}
\def\({\left(}
\def\){\right)}
\def\[{\left[}
\def\]{\right]}

\def\Journal#1#2#3#4{{#1} {\bf #2}, #3 (#4)}
\def\RPP{{Rep. Prog. Phys.}}
\def\PRC{{Phys. Rev. C.}}
\def\PRD{{Phys. Rev. D.}}
\def\FP{{Foundations of Physics}}
\def\ZPA{{Z. Phys.A.}}
\def\NPA{{Nucl. Phys. A.}}
\def\JPG{{J. Phys. G Nucl. Part}}
\def\PRL{{Phys. Rev. Lett.}}
\def\PRpt{{Phys. Rep.}}
\def\PLB{{Phys. Lett. B}}
\def\AP{{Ann. Phys (N.Y.)}}
\def\EPJA{{Eur. Phys. J.A}}
\def\NP{{Nucl. Phys}}
\def\ZP{{Z. Phys}}
\def\RMP{{Rev. Mod. Phys}}
\def\IJMPE{{Int. J. Mod. Phys. E}}
\input epsf

\title{Neutrino Electromagnetic Form Factors Effect on the Neutrino Cross Section 
in Dense Matter}

\author{A. Sulaksono, C.K. Williams, P.T.P. Hutauruk, T. Mart}

\affiliation{Departemen Fisika, FMIPA, Universitas Indonesia,
Depok, 16424, Indonesia}

\begin{abstract}
The sensitivity of the differential cross section of the interaction between neutrino-electron with dense matter to the possibly nonzero neutrino electromagnetic properties has been investigated. Here, the relativistic mean field model inspired by effective field theory has been used to describe non strange dense matter, both with and without the neutrino trapping. We have found that the cross section becomes more sensitive to the constituent distribution of the matter, once electromagnetic properties of the neutrino are taken into account. The effects of electromagnetic properties of neutrino on the cross section become more significant for the neutrino magnetic moment $\mu_{\nu}$ $>$ $10^{-10}\mu_B$ and for the neutrino charge radius $R$   $>$ $10^{-5} {\rm MeV}^{-1}$. 
\end{abstract}


\maketitle

\newpage

\section{Introduction}
\label{sec_intro}
The demand for precise  information on the neutrino transport in the investigations of astrophysical phenomena, such as supernovae explosion and the structure of protoneutron stars, has stimulated several studies on neutrino interactions in matter at high densities~\cite{reddy1,niembro01,parada04,Horo2,mornas1,mornas2,reddy2,horowitz91,yama,caiwan,margue,chandra,cowel,caroline05}. Even recently, a realistic neutrino opacity in neutron rich matter for the supernovae simulation which takes into account the correlations and weak magnetism of nucleons in the finite temperature calculations has been also considered \cite{Horo2}.

In the standard model, massless neutrinos have zero magnetic moment and electronic charge. However, there are evidences that the laboratory bounds on the neutrino-electron magnetic moment ($\mu_{\nu}$) is smaller than $1.0 \times 10^{-10} \mu_B$ at the 90\% confidence level~\cite{munu}, where $\mu_B$ is the Bohr magneton. Stronger bound  of $\mu_{\nu} \leq 3.0 \times 10^{-12} \mu_B$ also exists from astrophysical consideration, particularly from the study of the red giant population in the globular cluster~\cite{Raffelt99}. On the other hand the charge radius of $\nu_e$ has been bounded by the LAMF experiment~\cite{vilain} to be $R^2=(0.9 \pm 2.7)\times 10^{-32}~{\rm cm}^{2}=(22.5 \pm 67.5)\times 10^{-12}~{\rm MeV}^{-2}$, while the plasmon decay in the globular cluster star predicts the limit of $e_{\nu} \leq 2 \times 10^{-14}\,e$~\cite{Raffelt99}, where $e$ is electron charge. Recent discussions on the effects of the neutrino electromagnetic properties in astrophysics can be seen in, e.g., Ref.~\cite{Raffelt99}, while for the case of the solar neutrino problem one can consult, e.g., Ref.~\cite{friedland}.

So far, there has been no calculation of the interaction of neutrinos with dense matter  which considers neutrino electromagnetic form factors. To this end we extend our previous report of  the interaction of neutrinos with electrons gas~\cite{caroline05} to the interaction of neutrinos with the non strange dense stellar matter which takes into account the effects of the  trapped neutrinos in matter, but still use the zero temperature approximation. The validity of this approximation is fulfilled by the fact that temperature effects on the equation of state (EOS) of supernovae matter and maximum masses of protoneutron stars are smaller than those without neutrino trapping~\cite{hua03}. We note that considerable efforts have been devoted to study the effective electromagnetic coupling of neutrinos in the thermal background of particles~\cite{nives1,nives2,sawyer,samina}. However, the particles under investigation so far are electron and nucleon gasses. Clearly, the effect of nucleon correlations has not yet been studied. References~\cite{nives1,samina} have shown that the densities and temperature yield enhancements to the form factor with standard charge radius, but they do not give a significant effect on the electric and magnetic dipole form factors. Moreover, they can only give a significant contribution if $T/m$ and $\mu/m$ are large, where $T$, $\mu$ and $m$ are the temperature, chemical potential and mass of the corresponding particles in the thermal bath, respectively~\cite{samina}. Therefore, the expectation that the temperature and hadron correlations can strongly reduce the electromagnetic properties of neutrinos seems to be unlikely. For simplicity, the RPA correlations are still neglected. The importance of the neutrino trapping in the supernova dynamical evolution has been, for example, pointed out in Ref.~\cite{chiapparini}. We choose this kind of matter, because we suspect that the effects of the neutrino electromagnetic properties would be more pronounced in the protons- and electrons rich matter.  Different from Ref.~\cite{chiapparini}, which used the standard relativistic mean field (RMF) model to calculate the EOS, in this work we use the RMF model inspired by the effective field theory (E-RMF model)~\cite{Furnstahl96}.

The construction of this paper is as follows. Section~\ref{sec_nucmod} consists of a brief review of the matter models used in this work. The analytical results of the neutrino electromagnetic form factors effects on the cross section are given in section~\ref{sec_EFFN}. In section~\ref{sec_rslt}, numerical results and their discussions are presented. Finally, the conclusion is given in section~\ref{sec_sum}.
  
\section{Matter Models}
\label{sec_nucmod}
To describe nucleons interactions, we use the effective Lagrangian density given in Refs.~\cite{Furnstahl96, Wang00}. The wide range applications of this model have been discussed in detail in Refs.~\cite{sil,arumu}. In our work, leptons are assumed to be free (Fermi gas).  The framework of our calculation is the relativistic mean field approximation, which means that we treat the nucleon interactions self consistently. The explicit form of the Lagrangian densities can be seen in appendix~\ref{sec_app}. By solving the  Euler-Lagrange equations for the Lagrangian densities given in appendix~\ref{sec_app}, the equations of state for nucleons, mesons and leptons can be obtained, and then, can be used to calculate all matter properties required in this work. The following constraints are used to calculate the fraction of every constituent in matter:
\begin{itemize}
\item balance equation for chemical potentials
\be
\mu_{n}+\mu_{\nu_e} = \mu_{p}+\mu_{e},
\ee
\item conservation of charge neutrality
\be
\rho_e +\rho_{\mu}=\rho_p,
\ee
\end{itemize}
where the total density of baryon is given by
\be
\rho_B =\rho_n+\rho_p,
\ee
while the fixed electronic-leptonic fraction is defined as
\be
Y_{l_e}=\frac{\rho_e+\rho_{\nu_e}}{\rho_B}\equiv Y_e+Y_{\nu_e}.
\ee
Coupling constants for all parameter sets used in this work are shown in Table~\ref{tab:params1}.

\begin{table}
\centering
\caption {Numerical values of coupling constants used in the parameter sets.}\label{tab:params1}
\begin{tabular}{crcr}
\hline\hline Parameter &G2~~ & NL-3~~& G2* \\\hline
$m_S/M$        &0.554~~& 0.541~~& 0.554 \\
$g_S/(4 \pi)$ &0.835~~& 0.813~~& 0.835\\
$g_V/(4 \pi)$ &1.016~~& 1.024~~& 1.016\\
$g_R/(4 \pi)$ &0.755~~& 0.712~~& 0.938\\
$\kappa_3$     &3.247~~& 1.465~~& 3.247\\
$\kappa_4$     &0.632~~&$-5.668$~~~~&0.632\\
$\zeta_0$      &2.642~~& 0~~& 2.642\\
$\eta_1$       &0.650~~& 0~~& 0.650\\
$\eta_2$       &0.110~~& 0~~& 0.110\\
$\eta_{\rho}$  &0.390~~& 0~~& 4.490\\\hline\hline
\end{tabular}\\
\end{table}
\section{Effects of  the Neutrino Electromagnetic Form Factors }
\label{sec_EFFN}
In this section we calculate the neutrino-matter cross section, in which the electromagnetic form factors of the neutrino-electron and the weak magnetism of nucleons are explicitly taken into account. We start with the Lagrangian density of  neutrino matter interactions for each constituent in the form of 
\bea
{\mathcal L}_{\rm int}^{ j}&=&\frac{G_F}{\sqrt{2}}
(\bar{\nu}\Gamma_{\rm W}^{\mu} \nu)(\bar{\psi} J_{\mu}^{{\rm W}~{ j}} \psi)
+\frac{4 \pi \alpha}{q^2}(\bar{\nu}\Gamma_{\rm EM}^{\mu} \nu)(\bar{\psi} J_{\mu}^{{\rm EM}~{ j}} \psi),
\label{eq:lagden1}
\eea
where $G_F$ and $\alpha$ are the coupling constant of weak interaction and the electromagnetic fine structure constant, respectively, and ${ j}$= $n, p, e^{-}, {\mu}^{-}$. The parity violating vertex of neutrino is given by
\begin{equation}
\Gamma_{\rm W}^{\mu}=\gamma^\mu(1-\gamma^5),
\end{equation}
while the electromagnetic properties of Dirac neutrinos are described in 
terms of four form factors, i.e., $f_{1\nu}, g_{1\nu}, f_{2\nu}$ and 
$g_{2\nu}$, which stand for the Dirac, anapole, magnetic, and electric 
form factors, respectively. The electromagnetic vertex $\Gamma_{\rm EM}^{\mu}$ contains electromagnetic form factors~\cite{Kerimov, Mourao}. Explicitly, it reads
\bea
\Gamma_{\rm EM}^{\mu}=f_{m\nu}\gamma^{\mu}
+ g_{1\nu}\gamma^{\mu}\gamma^{5}-(f_{2\nu}+ig_{2\nu}\gamma^5)
\frac{P^\mu}{2m_e}.
\eea
where $f_{m\nu}=f_{1\nu}+(m_\nu / m_e)f_{2\nu}$, 
$P^{\mu} = k^{\mu} +k^{\mu\prime}$, $m_\nu$ and $m_e$ are the neutrino and 
electron masses, respectively.
In the static limit, the reduced Dirac form factor $f_{1\nu}$ and the 
neutrino anapole form factor $g_{1\nu}$ are related to the vector and 
axial vector charge radii $\langle{R^2_V}\rangle$ and $\langle{R^2_A}\rangle$ 
through~\cite{Kerimov} 
\begin{equation}
f_{1\nu}(q^2)=\frac{1}{6} \langle{R^2_V}\rangle q^2 \quad\textrm{and}\quad 
g_{1\nu}(q^2)=\frac{1}{6}\langle{R^2_A}\rangle q^2.
\label{effac}
\end{equation}
where the neutrino charge radius is defined by $R^2$ = $\langle{R^2_V}\rangle$ + $\langle{R^2_A}\rangle$. In the limit of $q^2\to0$, $f_{2\nu}$ and $g_{2\nu}$,  respectively,  define
the neutrino magnetic moment  and the (CP violating) electric dipole moment~\cite{nardi,Kerimov}, i.e.,
\begin{equation}
\mu_{\nu}^{m}=f_{2\nu}(0)\mu_B \quad\textrm{and}\quad \mu_{\nu}^e=g_{2\nu}(0)\mu_B,
\label{mffac}
\end{equation}
where  $\mu_{\nu}^2$=${(\mu_{\nu}^m)}^{2}$+ ${(\mu_{\nu}^e)}^{2}$. The explicit form of $J_\mu^{{\rm W}~j}$~\cite{Horo2} is given by
\be
J_\mu^{{\rm W}~j}=F_1^{{\rm W}~j}\gamma_\mu-G_A^{j}\gamma_\mu \gamma^5
+i F_2^{{\rm W}~j} ~\frac{\sigma_{\mu \nu}q^{\nu}}{2~M},
\ee
and for $J_\mu^{{\rm EM}~{ j}}$~\cite{Vogel} it reads 
\be
J_\mu^{{\rm EM}~{ j}}=F_1^{{\rm EM}~{ j}}\gamma_\mu+i F_2^{{\rm EM}~{ j}} ~\frac{\sigma_{\mu \nu}q^{\nu}}{2~M}.
\ee
\begin{table}
\centering
\caption {Weak form factors in the limit  of $q^2\to0$. Here we use $\sin^2\theta_w=0.231$, $g_A= 1.260$, $\mu_p=1.793$ and $\mu_n=-1.913$~\cite{Horo2}.}\label{tab:copconst0}
\begin{tabular}{ccrc}
\hline\hline Target & $F_1^{\rm W}$ & $G_A$~~~~ & $F_2^{\rm W}$ \\\hline\hline
$n$     &$-0.5$~~  & $- g_A/2$~~  &$-1/2(\mu_p-\mu_n)-2\sin^2\theta_w\mu_n$\\
$p$     &$0.5-2\sin^2\theta_w$  & $g_A/2$~~ &~~$1/2(\mu_p-\mu_n)-2\sin^2\theta_w\mu_p$\\
$e$     &$0.5+2\sin^2\theta_w$  & 1/2~~ &0\\
$\mu$ &$-0.5+2\sin^2\theta_w$~~  & $-1/2$~~  &0 \\\hline\hline
\end{tabular}\\
\end{table}

\begin{table}
\centering
\caption {Electromagnetic form factors in the limit  of $q^2\to0$~\cite{Vogel}.}\label{tab:copconst1}
\begin{tabular}{ccc}
\hline\hline ~~Target~~ &~~~~~~ $F_1^{\rm EM}$~~~~~~ & ~~$F_2^{\rm EM}$~~ \\\hline\hline
$n$      &0   & $\mu_n$   \\
$p$     &1  &  $\mu_p$  \\
$e$     &1  & 0   \\
$\mu$  &1  & 0 \\\hline\hline
\end{tabular}\\
\end{table}
Using the Lagrangian density given by Eq.\,(\ref{eq:lagden1}), we can now calculate the differential cross section. Using the standard method we obtain
\begin{eqnarray}
\Biggl(\frac{1}{V}\frac{d^3\sigma}{d^2\Omega'dE_\nu'}\Biggr)&=& 
-\frac{1}{16\pi^2}\frac{E_\nu'}{E_\nu}\left[ \Biggl( \frac{G_F}{\sqrt{2}} 
\right)^2 {(L^{\mu\nu}_\nu\Pi^{\rm Im}_{\mu\nu})}^{\rm (W)}\nonumber\\
&+& \left( \frac{4\pi\alpha}{q^2}\right) ^2 {(L^{\mu\nu}_\nu\Pi^{\rm Im}_{\mu\nu})}^{\rm (EM)}+ 
\frac{8G_F\pi\alpha}{q^2\sqrt{2}} {(L^{\mu\nu}_{\nu}\Pi^{\rm Im}_{\mu\nu})}^{\rm (INT)} \Biggr].
\label{1}
\end{eqnarray}
Here $E_{{\nu}}$ and $ E'_{{\nu}}$ are the initial and final neutrino energies, respectively,  $G_{F}= 1.023{\times} 10^{-5}/M^{2}$ is the weak coupling, and $M$ is the nucleon mass. The neutrino tensors for the weak contribution is given by 
\begin{eqnarray}
L^{\mu\nu {\rm (W)}}_{\nu}  &=& 8\left[2k^{\mu}k^{\nu}-(k^{\mu}q^{\nu}+k^{\nu}q^{\mu})+g^{\mu\nu}(k \cdot q)-i\epsilon^{\alpha\mu\beta\nu}k_{\alpha}k_{\beta}'\right],
\end{eqnarray}
while for the electromagnetic contribution,
\begin{eqnarray}
L^{\mu\nu {\rm (EM)}}_{\nu} &=& 4(f^{2}_{m\nu}+g^{2}_{1\nu})[2k^{\mu}k^{\nu}-(k^{\mu}q^{\nu}+k^{\nu}q^{\mu})
+ g^{\mu\nu} (k \cdot q)]\nonumber\\&-&8if_{m\nu}g_{1\nu}\epsilon^{\alpha\mu\beta\nu}(k_{\alpha}k_{\beta}')\nonumber\\
&-&\frac{f^{2}_{2\nu}+g^{2}_{2\nu}}{m_e^2}(k \cdot q) 
 [4k^{\mu}k^{\nu}-2 (k^{\mu}q^{\nu}+q^{\mu}k^{\nu})+q^{\mu}q^{\nu}],
\end{eqnarray}
 and for the interference contribution,
\begin{eqnarray}
L^{\mu\nu {\rm (INT)}}_{\nu} &=& 4(f_{m\nu}+g_{1\nu})[2k^{\mu}k^{\nu}-(k^{\mu}q^{\nu}+k^{\nu}q^{\mu})
+ g^{\mu\nu}(k \cdot q)-i\epsilon^{\alpha\mu\beta\nu}k_{\alpha}k_{\beta}'], 
\end{eqnarray}
with $k$ the initial neutrino four-momentum and $q=(q_{0},{\vec{q}})$ the four-momentum transfer. The polarization tensors ${\Pi}^{{\mu}{\nu}}$ for the weak (W), electromagnetic (EM) and interference (INT) terms, which define the target particles, can be written as 
\begin{eqnarray}
{\Pi}^{ j}_{{\mu}{\nu}}(q) &=& -i \int\frac{d^{4}p}{(2{\pi})^{4}}{\rm Tr}[G^{ j}(p)J^{ j}_{{\mu}}G^{ j}(p+q)J^{ j}_{{\nu}}],
\label{polar}
\end{eqnarray}
where $p=(p_0,{\vec{p}})$ is the corresponding initial four-momentum and $G(p)$ is the target particle propagator. The explicit form for nucleons is given by
\bea
G^{n,p}(p)&=&(\pslash^*+M^*)~\Biggr[\frac{1}{p^*2-M^{*~2}+i\epsilon}+\frac{ i~\pi}{E^*}\,\delta(p_0^*-E^*)\,\theta(p_F^{p,n}-\mid\vec{p}\mid)\Biggr],
\label{propagator}
\eea
where $E^*$ = $E$ + $\Sigma_0$ indicates the nucleon effective energy,  and $M^*$ = $E$ + $\Sigma_S$ is the nucleon effective mass. $\Sigma_0$ and $\Sigma_S$ are the scalar and time like self energies, respectively. The lepton propagators have similar expressions, only the starred quantities in Eq.\,(\ref{propagator}) are replaced by the free ones. Explicitly, Eq.\,(\ref{polar}) for each constituent can be written as
\bea
{\Pi}^{{\rm Im (W)} j}_{{\mu}{\nu}}&=& (F_1^{{\rm W}{ j}~2}+G_A^{{ j}~2}){\Pi}^{{\rm V} j}_{{\mu}{\nu}}\nonumber\\&+&\Biggr(G_A^{{ j}~2}+\frac{q^2}{2 m M} F_1^{{\rm W}{ j}} F_2^{{\rm W}{ j}}\Biggr){\Pi}^{{\rm A} j} g_{{\mu}{\nu}}\nonumber\\&-&2~\Biggr( F_1^{{\rm W}{ j}} G_A^{ j}+\frac{m}{M}F_2^{{\rm W}{ j}} G_A^{ j}\Biggr){\Pi}^{{\rm V-A} j}_{{\mu}{\nu}}\nonumber\\&+&\frac{F_2^{{\rm W}{ j}~2}}{M^2}\Biggr[(m^2+\frac{q^2}{4})(q^2  g_{{\mu}{\nu}}-q_{\mu}q_{\nu})-\frac{q^2}{8}{\Pi}^{{\rm V} j}_{{\mu}{\nu}}\Biggr],
\eea
\bea
{\Pi}^{{\rm Im (EM)} j}_{{\mu}{\nu}}&=& F_1^{{\rm EM}{ j}~2}{\Pi}^{{\rm V} j}_{{\mu}{\nu}}\nonumber\\&+&\frac{q^2}{2 m M} F_1^{{\rm EM}{ j}} F_2^{{\rm EM}{ j}} {\Pi}^{{\rm A} j} g_{{\mu}{\nu}}\nonumber\\&+&\frac{F_2^{{\rm EM}{ j}~2}}{M^2}\Biggr[(m^2+\frac{q^2}{4})(q^2  g_{{\mu}{\nu}}-q_{\mu}q_{\nu})-\frac{q^2}{8}{\Pi}^{{\rm V} j}_{{\mu}{\nu}}\Biggr],
\eea
\bea
{\Pi}^{{\rm Im (INT)} j}_{{\mu}{\nu}}&=& ( F_1^{{\rm W}{ j}} F_1^{{\rm EM}{ j}}+\frac{q^2}{4 M^2} F_2^{{\rm W}{ j}} F_2^{{\rm EM}{ j}}){\Pi}^{{\rm V} j}_{{\mu}{\nu}}\nonumber\\&+&\Biggr[\frac{F_2^{{\rm W}{ j}} F_2^{{\rm EM}{ j}}}{4 M^2}(1+\frac{q^2}{4 m^2})-\frac{(F_1^{{\rm W}{ j}} F_2^{{\rm EM}{ j}}+F_2^{{\rm W}{ j}} F_1^{{\rm EM}{ j}})}{4 m M}\Biggr]\nonumber\\&\times& (q^2  g_{{\mu}{\nu}}-q_{\mu}q_{\nu}){\Pi}^{{\rm A} j}+\Biggr(\frac{m}{M}F_2^{{\rm EM}{ j}} G_A^{ j}-F_1^{{\rm EM}{ j}} G_A^{ j}\Biggr){\Pi}^{{\rm V-A} j}_{{\mu}{\nu}},
\eea
where for $j=n,p$ (nucleons), $m$ is equal to $M^*$ and $M$ is the nucleon mass, while for  $j=e^-,\mu^-$ (leptons), $m$ is equal to $M$, the lepton mass.

Due to the current conservation and translational invariance, the vector 
polarization $\Pi^{{\rm Im}{V}}_{\mu\nu}$ of every contribution consists of two independent 
components which we choose to be in the frame of 
$q^\mu \equiv(q_0,\vert{\vec{q}}\vert,0,0)$, i.e., 
\begin{eqnarray}
\Pi_{T}&=&\Pi^V_{22}=\Pi^V_{33} \quad\textrm{and}\nonumber\\
\Pi_{L}&=&-(q^2_{\mu}/\vert\vec{q}\vert^2)\Pi^V_{00}.\nonumber 
\end{eqnarray}
The axial-vector and the mixed pieces are found to be
\begin{eqnarray}
\Pi^{\rm Im (V-A)}_{\mu\nu}(q)=i\epsilon_{\alpha\mu 0\nu}q_{\alpha}\Pi_{VA},
\end{eqnarray}

The explicit forms of $\Pi_{T}$, $\Pi_{L}$, $\Pi_{VA}$ and $\Pi_A$ for nucleons
are~ \cite{Horowitz01}
\bea
\Pi_T &=&\frac{1}{4\pi\vert\vec{q}\vert} 
\Biggr[ \(M^{*~2} + \frac{q^2}{4\vert\vec{q}\vert^2} + \frac{q^2}{2}\) 
(E_F-E^*) 
\nonumber\\ &+& \frac {q_0 ~q^2} {2\vert\vec{q}\vert} (E^2_F-E^{* 2})
+ \frac{q^2}{3\vert\vec{q}\vert} (E^3_F-E^{*3})\Biggr],
\label{PiT}
\eea
\be
\Pi_L =  \frac{q^2}{2\pi\vert\vec{q}\vert^3}\[\frac{1}{4} (E_F-E^*) +\frac{q_0}{2} (E^2_F-E^{*2}) + \frac{1}{3}(E^3_F-E^{*3})\],
\label{PiL}
\ee
\begin{eqnarray}
\Pi_{VA}=\frac{iq^2}{8\pi\vert\vec{q}\vert^3}[(E^2_F-E^{* 2}) +q_0 (E_F-E^*)]. 
\label{PiVA}
\end{eqnarray}
\begin{eqnarray}
\Pi_A=\frac{i}{2\pi\vert\vec{q}\vert}M^{* 2}(E_F-E^*).
\label{PiA}
\end{eqnarray}
For leptons, they have also similar expressions, only the starred quantities in Eqs.~(\ref{PiT}),~(\ref{PiL}),~(\ref{PiVA}) and~(\ref{PiA}) are replaced by the free ones. Thus, the  analytical form of Eq.\,(\ref{1}) can be obtained from the 
contraction of every polarization and neutrino tensors couple ($L^{\mu\nu }\Pi_{\mu\nu}$) mentioned previously. The results are
\bea
{(L^{\mu\nu}_\nu\Pi^{\rm Im }_{\mu\nu})}^{\rm (W)}&=&-8~ q^2\sum_{{ j}=n, p, e^{-}, {\mu}^{-}}\Biggr[ A_{\rm W}^{ j} ( \Pi_L^{ j} + \Pi_T^{j} )+ B_{\rm 1~W}^{ j} \Pi_T^{ j} + B_{\rm 2~W}^{ j} \Pi_A^{ j}+ C_{\rm W}^{ j}\Pi_{VA}^{ j} \Biggr],~~~
\label{ctrweak} \\
{(L^{\mu\nu}_\nu\Pi^{\rm Im }_{\mu\nu})}^{\rm (EM)}&=& q^2\sum_{{ j}=n, p, e^{-}, {\mu}^{-}}\Biggr[ A_{\rm EM}^{ j}( \Pi_L^{ j} + \Pi_T^{ j} )+ B_{\rm 1~EM}^{ j} \Pi_T^{ j} + B_{\rm 2~EM}^{ j} \Pi_A^{ j} \Biggr],~~~ 
\label{ctrem}\\
{(L^{\mu\nu}_\nu\Pi^{\rm Im }_{\mu\nu})}^{\rm (INT)}&=&-4 ~q^2\sum_{{ j}=n, p, e^{-}, {\mu}^{-}}\Biggr[ A_{\rm INT}^{ j}( \Pi_L^{ j} + \Pi_T^{ j} )+ B_{\rm 1~INT}^{ j} \Pi_T^{ j} + B_{\rm 2~INT}^{ j} \Pi_A^{ j}+ C_{\rm INT}^{ j}\Pi_{VA}^{ j} \Biggr], \nonumber\\
\label{ctrint}
\eea  
where the functions in the front of every polarization terms of Eqs.~(\ref{ctrweak}) -~(\ref{ctrint}) are given by
\bea
 A_{\rm W}^{ j}&=&\(\frac{2 E(E-q_0)+\frac{1}{2} q^2}{{\vert\vec{q}\vert}^2}\)\Biggr[F_1^{{\rm W}{ j}~2}+G_A^{{ j}~2}-\frac{F_2^{{\rm W}{ j}~2}q^2}{4 M^2}\Biggr],\nonumber\\B_{\rm 1~W}^{ j}&=&\Biggr[F_1^{{\rm W}{ j}~2}+G_A^{{ j}~2}-\frac{F_2^{{\rm W}{ j}~2}q^2}{4 M^2}\Biggr],\nonumber\\B_{\rm 2~W}^{ j}&=&-\Biggr[G_A^{{ j}~2}+\frac{q^2}{2 m M} F_1^{{\rm W}{ j}} F_2^{{\rm W}{ j}}-\frac{F_2^{{\rm W}{ j}~2}q^2}{4 M^2}(1+\frac{q^2}{4 m^2})\Biggr],\nonumber\\ C_{\rm W}^{ j}&=&-2~(2E-q_0)~\Biggr[ F_1^{{\rm W}{ j}} G_A^{ j}+\frac{m}{M}F_2^{{\rm W}{ j}} G_A^{ j}\Biggr],
\eea
for the weak contributions, and
\bea
 A_{\rm EM}^{ j}&=&\Biggr[ \(\frac{2 E(E-q_0)+\frac{1}{2} q^2}{{\vert\vec{q}\vert}^2}\)(b ~q^2-a)+\frac{1}{2} b ~q^2\Biggr]\Biggr[F_1^{{\rm EM}{ j}~2}-\frac{F_2^{{\rm EM}{ j}~2}q^2}{4 M^2}\Biggr],\nonumber\\B_{\rm 1~EM}^{ j}&=&-\frac{1}{2}(b ~q^2+a)\Biggr[F_1^{{\rm EM}{ j}~2}-\frac{F_2^{{\rm EM}{ j}~2}q^2}{4 M^2}\Biggr],\nonumber\\B_{\rm 2~EM}^{ j}&=&\frac{1}{2}(b ~q^2+a)\Biggr[\frac{q^2}{2 m M} F_1^{{\rm EM}{ j}} F_2^{{\rm EM}{ j}}-\frac{F_2^{{\rm EM}{ j}~2}q^2}{4 M^2}(1+\frac{q^2}{4 m^2})\Biggr],
\eea
for the electromagnetic contributions, and
\bea
 A_{\rm INT}^{ j}&=& c~ \(\frac{2 E(E-q_0)+\frac{1}{2} q^2}{{\vert\vec{q}\vert}^2}\)\Biggr[F_1^{{\rm W}{ j}}F_1^{{\rm EM}{ j}}+\frac{q^2}{4 M^2} F_2^{{\rm W}{ j}} F_2^{{\rm EM}{ j}}\Biggr],\nonumber\\B_{\rm 1~INT}^{ j}&=& c~\Biggr[F_1^{{\rm W}{ j}}F_1^{{\rm EM}{ j}}+\frac{q^2}{4 M^2} F_2^{{\rm W}{ j}} F_2^{{\rm EM}{ j}}\Biggr],\nonumber\\B_{\rm 2~INT}^{ j}&=&-c~q^2\Biggr[\frac{F_2^{{\rm W}{ j}} F_2^{{\rm EM}{ j}}}{4 M^2}(1+\frac{q^2}{4 m^2})-\frac{(F_1^{{\rm W}{ j}} F_2^{{\rm EM}{ j}}+F_2^{{\rm W}{ j}} F_1^{{\rm EM}{ j}})}{4 m M}\Biggr],\nonumber\\ C_{\rm INT}^{ j}&=&c~(2E-q_0)~\Biggr[\frac{m}{M}F_2^{{\rm EM}{ j}} G_A^{ j}-F_1^{{\rm EM}{ j}} G_A^{ j}\Biggr],
\eea
for the interference contributions. The constants $a$, $b$ and $c$ are defined as 
\bea
a = 4 (f_{m \nu}^2+g_{1 \nu}^2),~ ~ ~ ~ ~
b = \frac{f_{2 \nu}^2+g_{2 \nu}^2}{m_e^2},~ ~ ~ ~ ~
c =f_{m \nu}+g_{1 \nu}.\nonumber
\eea
where these neutrino form factors, $f_{m \nu}$, $g_{1 \nu}$, $f_{2 \nu}$ and $g_{2 \nu}$, are related to the neutrino-electron dipole moment and charge radius through Eqs.~(\ref{effac}) and~(\ref{mffac}).
\section{Numerical Results and Discussions}
\label{sec_rslt}

\begin{figure*}[!t]
\centering
 \mbox{\epsfig{file=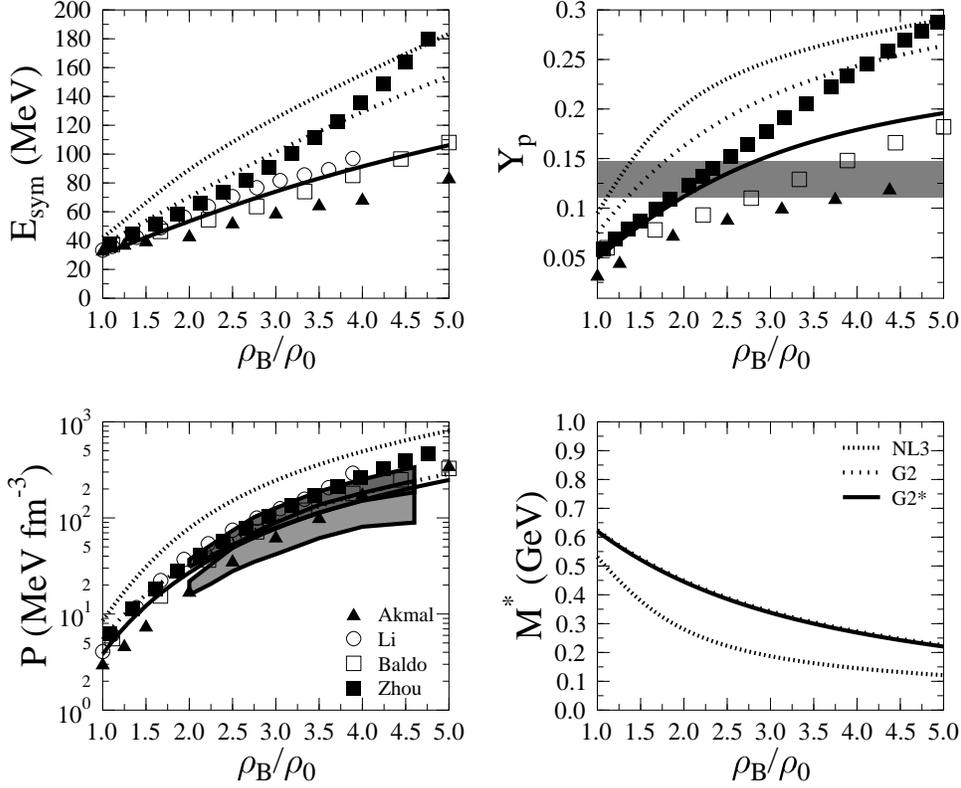,height=11.0cm}}
\caption{ Performance of the models. G2 is the  standard parameter set of the E-RMF model, G2* is  the parameter set of the E-RMF model with an adjusted isovector-vector channel (see appendix~\ref{sec_app} for details). NL3 is the parameter set of the standard RMF model. The symmetry energy $E_{\rm sym}$ of nuclear matter  is shown in the upper left panel. The pressure and $M^*$ of pure neutron matter (PNM) are given in the lower left and the lower right panels, respectively, and the neutron star proton fraction predictions can be seen in the upper right panel. Shaded region in the lower left panel corresponds to experimental data of Danielewicz  {\it et ~al.}~\cite{daniel02}, whereas shaded region in the upper right panel corresponds to the proton fraction  threshold for direct URCA process. For comparison, we also  show the results from variational calculation of Akmal {\it et ~al.}~\cite{akmal98}, Dirac Brueckner Hartree Fock (DBHF) calculation of Li {\it et ~al.}~\cite{li92}, Brueckner Hartree Fock (BHF) with AV14 potential plus the phenomenological 3BF of Baldo {\it et ~al.}~\cite{baldo97} and the recent Brueckner Hartree Fock (BHF) calculation with the meson-exchange microscopic model of Zhou {\it et ~al.}~\cite{Zhou04}.\label{G2*}}
\end{figure*}

\begin{figure*}
\centering
 \mbox{\epsfig{file=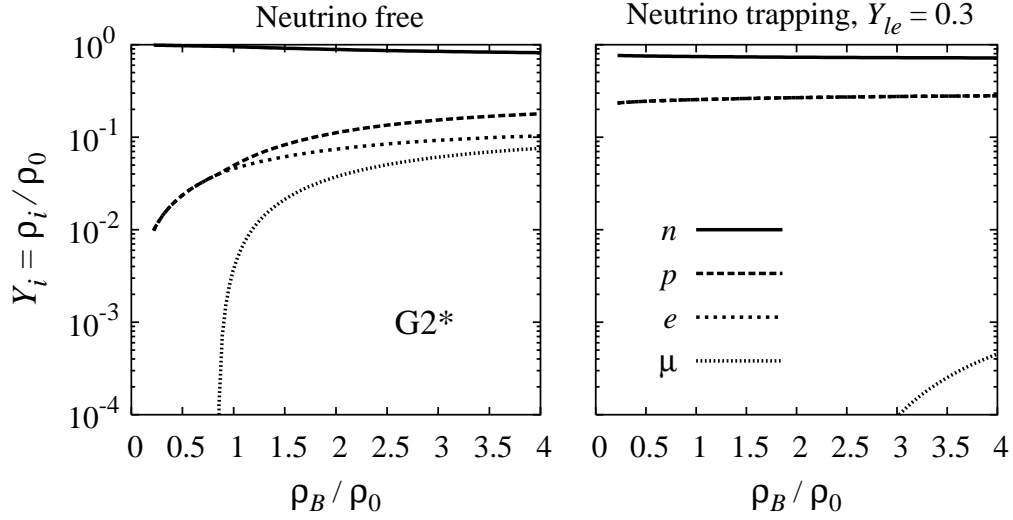, width=14cm}}
\caption{ Relative fraction of the individual constituent of the non strange matter as a function of  the ratio between baryon and saturation densities. Calculation has been performed by using the G2* parameter set. The case where neutrinos in matter are absent is shown in the left panel, while the case where neutrinos are trapped in matter with  a value of the electronic-leptonic fraction $Y_{l_e}=0.3$ is shown in the right panel.}
 \label{trap}
\end{figure*}

Before investigating the sensitivity of the neutrino-matter differential cross section to the neutrino electromagnetic form factors, we show the predictions of the nuclear model used  in the calculation (G2* parameter set) at high density in Figs.~\ref{G2*} and ~\ref{trap}. Figure~\ref{G2*} reveals that the G2* parameter set has the softest $E_{\rm sym}$ compared to the other parameter sets (G2 and NL3). As a consequence, it has the highest threshold density for the direct URCA process compared to the  other parameter sets. On the other hand, G2* and G2 have a similar trend in the  pure neutron matter (PNM) EOS and effective mass $M^*$, i.e., soft EOS and high value of $M^*$ at high densities. This fact indicates that the neutron star properties (masses, radii, etc) predicted by  G2* and G2  are quite similar. Comparisons between ERMF results and the Dirac Brueckner Hartree Fock (DBHF), Brueckner Hartree Fock (BHF) as well variational calculations are also shown in Fig.~\ref{G2*}, from which we can learn that G2* has an agreement in proton fraction ($Y_p$) with the recent BHF calculation of  Zhou {\it et ~al.}~\cite{Zhou04} at $\rho$ $\le ~2.5~\rho_B$, but at large densities,  their calculation predicts a larger  $Y_p$. On the other hand, and in general, Baldo {\it et ~al.}~\cite{baldo97} and  Akmal {\it et ~al.}~\cite{akmal98} obtained a relatively much smaller $Y_p$ than that of G2*. The $Y_p$ differences in all models originate from the differences in the predicted $E_{\rm sym}$.

\begin{figure*}
\centering
 \mbox{\epsfig{file=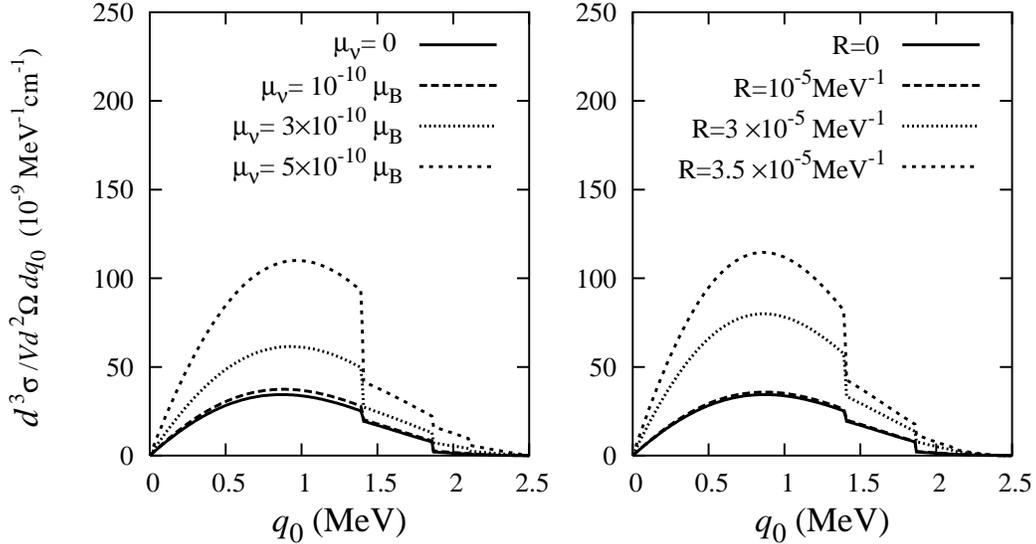, width=14cm}}
\caption{Total differential cross section as a function of $q_0$, calculated at fixed $q_1=2.5$ MeV, $E_{\nu}= 5$ MeV in neutrinoless matter. In the left panel  $R$ is fixed to zero and  $\mu_{\nu}$ is varied, whereas in the right panel  $\mu_{\nu}$ is fixed to zero and   $R$ is varied. }
 \label{crossna}
\end{figure*}

\begin{figure*}
\centering
 \mbox{\epsfig{file=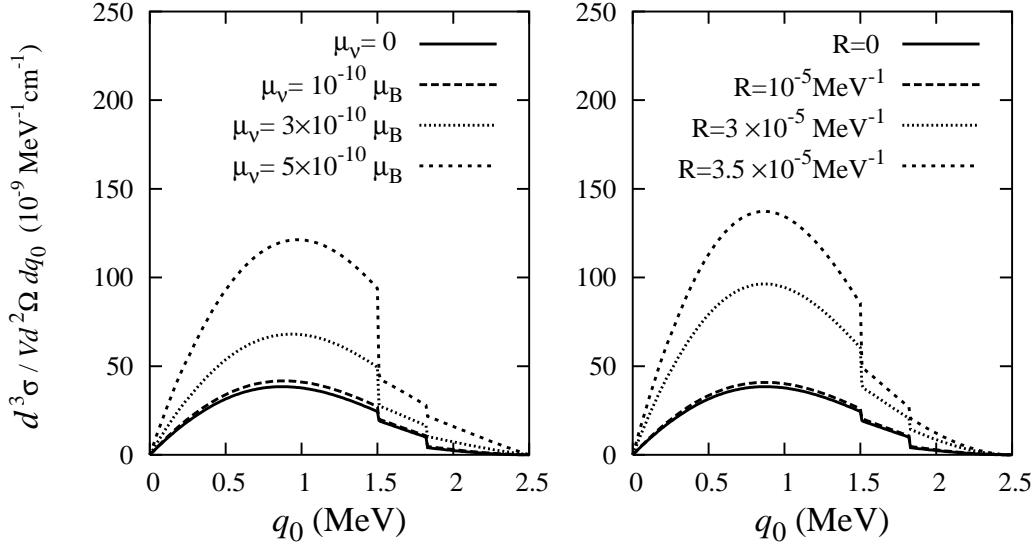, width=14cm}}
\caption{Same as in Fig.~\ref{crossna} but for the case where neutrinos are  trapped in matter with  $Y_{l_e}$ = 0.3. }
 \label{crossnt}
\end{figure*}

\begin{figure*}
\centering
 \mbox{\epsfig{file=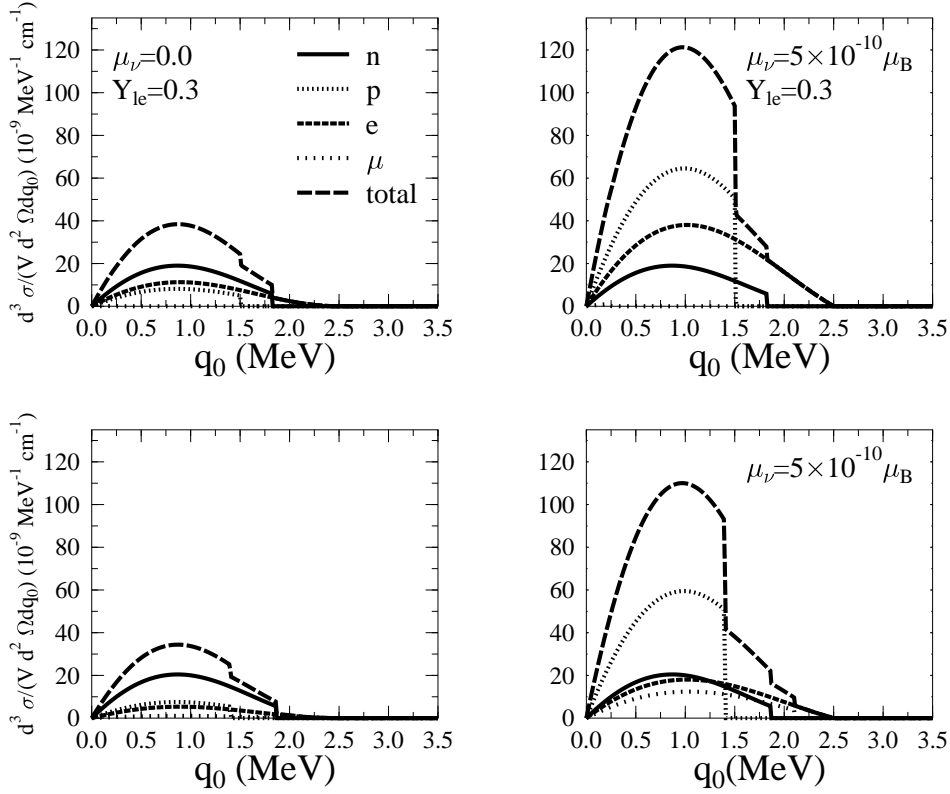, height=11.0cm}}
\caption{Total differential cross section and contributions from individual constituents as a function of $q_0$ obtained at fixed $q_1=2.5$ MeV, $E_{\nu}=5$ MeV, and $\rho_B= 2.5 \rho_0$, while the charge radius is set to $R=0$. The case where $\mu_{\nu}=0$ and neutrinos in matter are absent is given in the lower left panel, while for $\mu_{\nu}=0$ and $Y_{l_e}=0.3$  the result is given in the upper left panel. The case of $\mu_{\nu}= 5 \times 10^{-10}\mu_{B}$  and neutrinos are absent is given in the lower right panel, while for $\mu_{\nu}= 5\times 10^{-10}\mu_{B}$ and $Y_{l_e}=0.3$  the result is shown in the upper right panel.}
\label{totccc}
\end{figure*}

\begin{figure*}
\centering
 \mbox{\epsfig{file=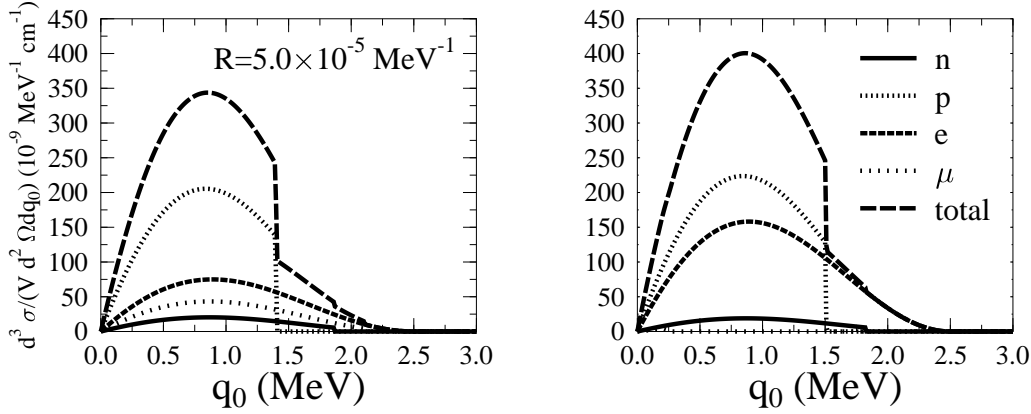, height=6cm}}
\caption{Total differential cross section and contributions from individual constituents as a function of $q_0$, calculated with fixed $q_1= 2.5$ MeV, $E_{\nu}=5$ MeV, and $\rho_B=2.5\rho_0$, while the neutrino magnetic moment ($\mu_{\nu}$) is set to zero and $R=5\times 10^{-5} {\rm MeV}^{-1}$. The case where neutrinos in matter are absent is given in the left panel, while the case where $Y_{l_e}=0.3$  is given in the right panel.}
\label{totc1}
\end{figure*}

\begin{figure*}
\centering
 \mbox{\epsfig{file=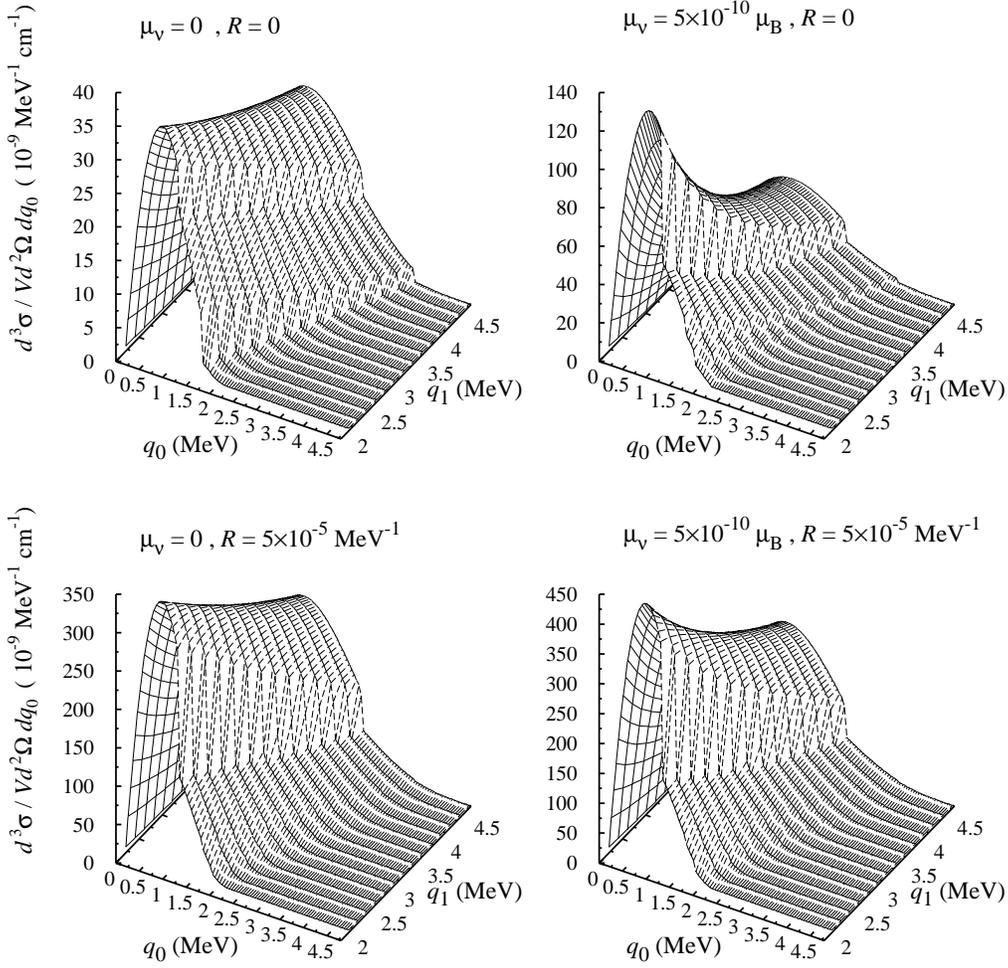,width=14.0cm}}
\caption{Total differential cross sections as functions of $q_0$ and  $q_1$  for the case where neutrinos are absent and  baryon density ($\rho_B$) is fixed to $2.5\rho_0$, while $E_{\nu}=5$ MeV.}
\label{cce5rb25}
\end{figure*}

\begin{figure*}
\centering
 \mbox{\epsfig{file=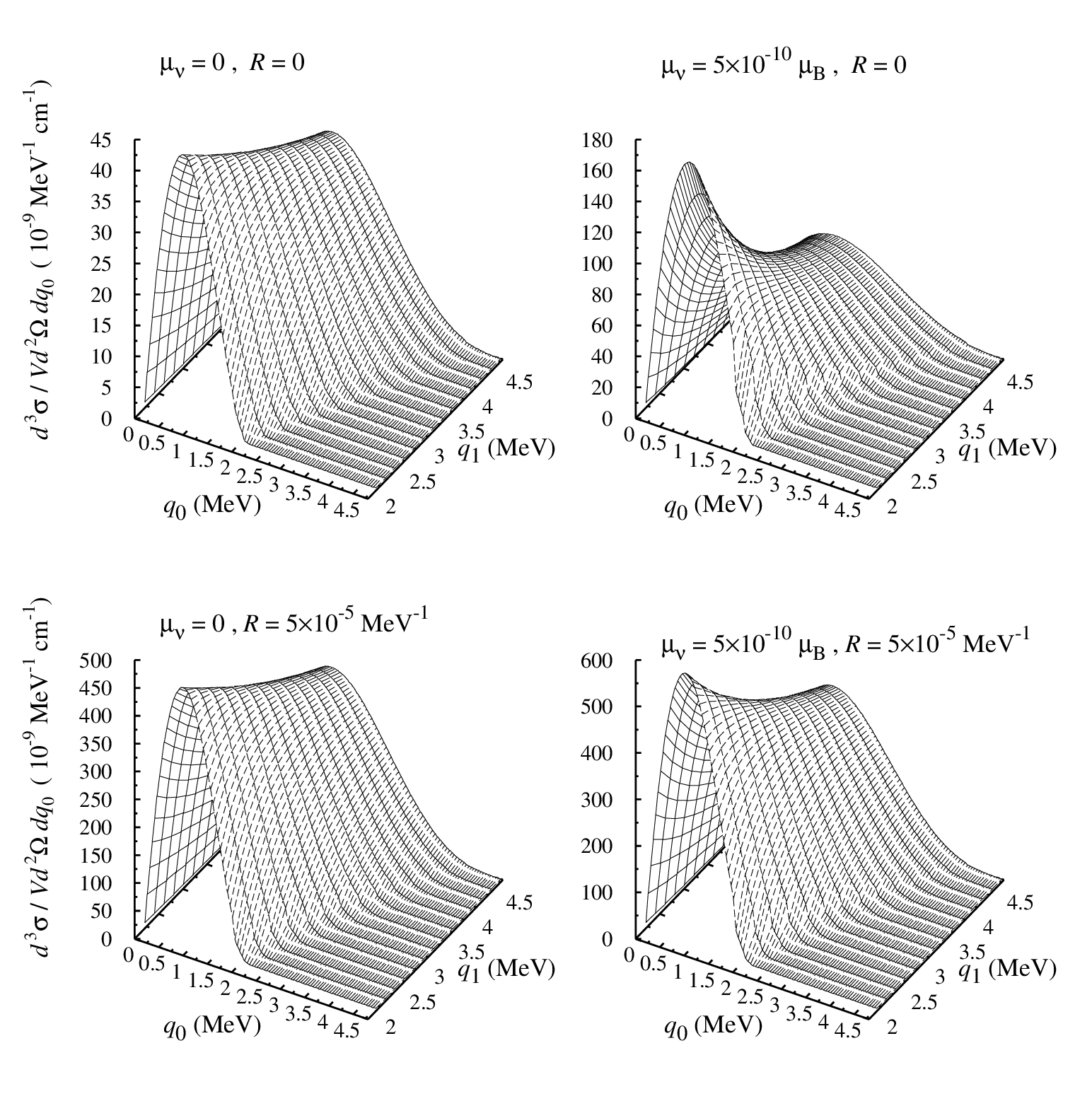,width=14.0cm}}
 \caption{Same as in Fig.~\ref{cce5rb25} but for  fixed baryon density $\rho_B=5\rho_0$.}
\label{cce5rb5}
\end{figure*}

\begin{figure*}
\centering
\mbox{\epsfig{file=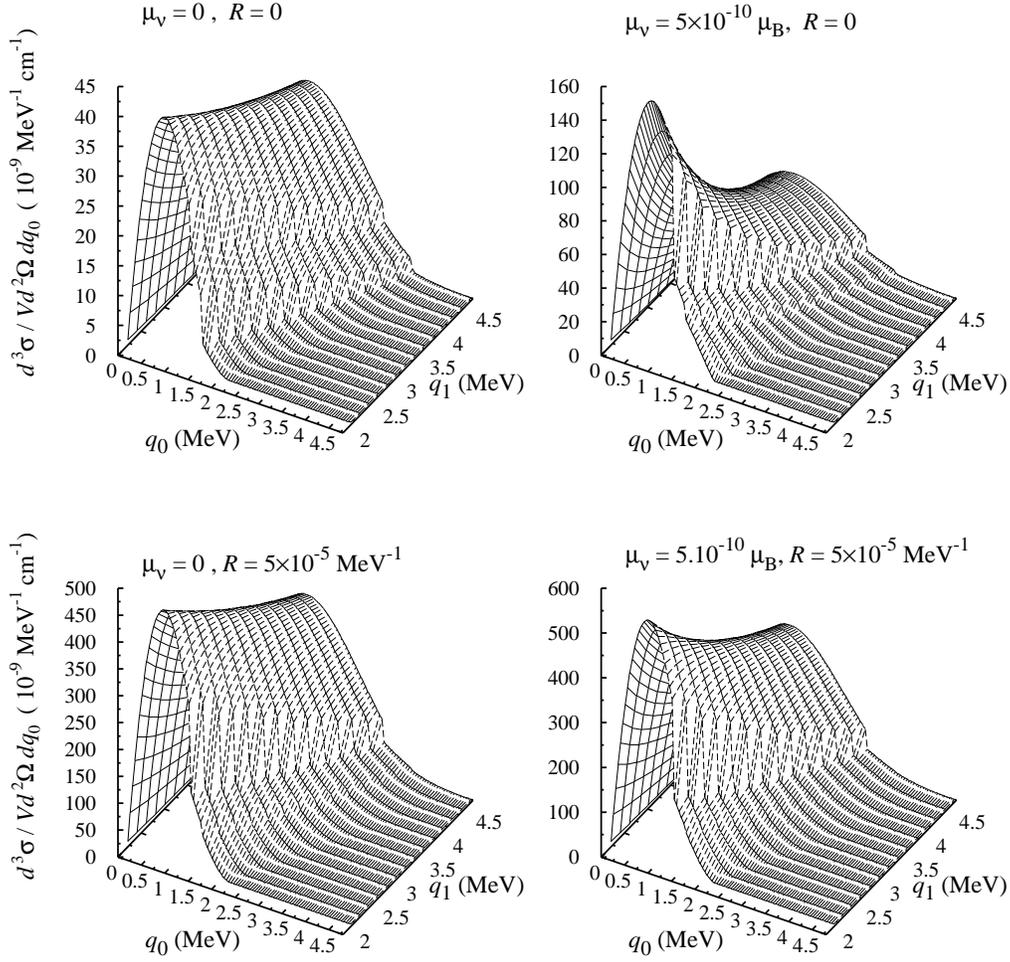, width=14.0cm}}
\caption{Same as in Fig.~\ref{cce5rb25} but for  $Y_{l_e}=0.3$.}
\label{ccpe5rb25}
\end{figure*}

\begin{figure*}
\centering
 \mbox{\epsfig{file=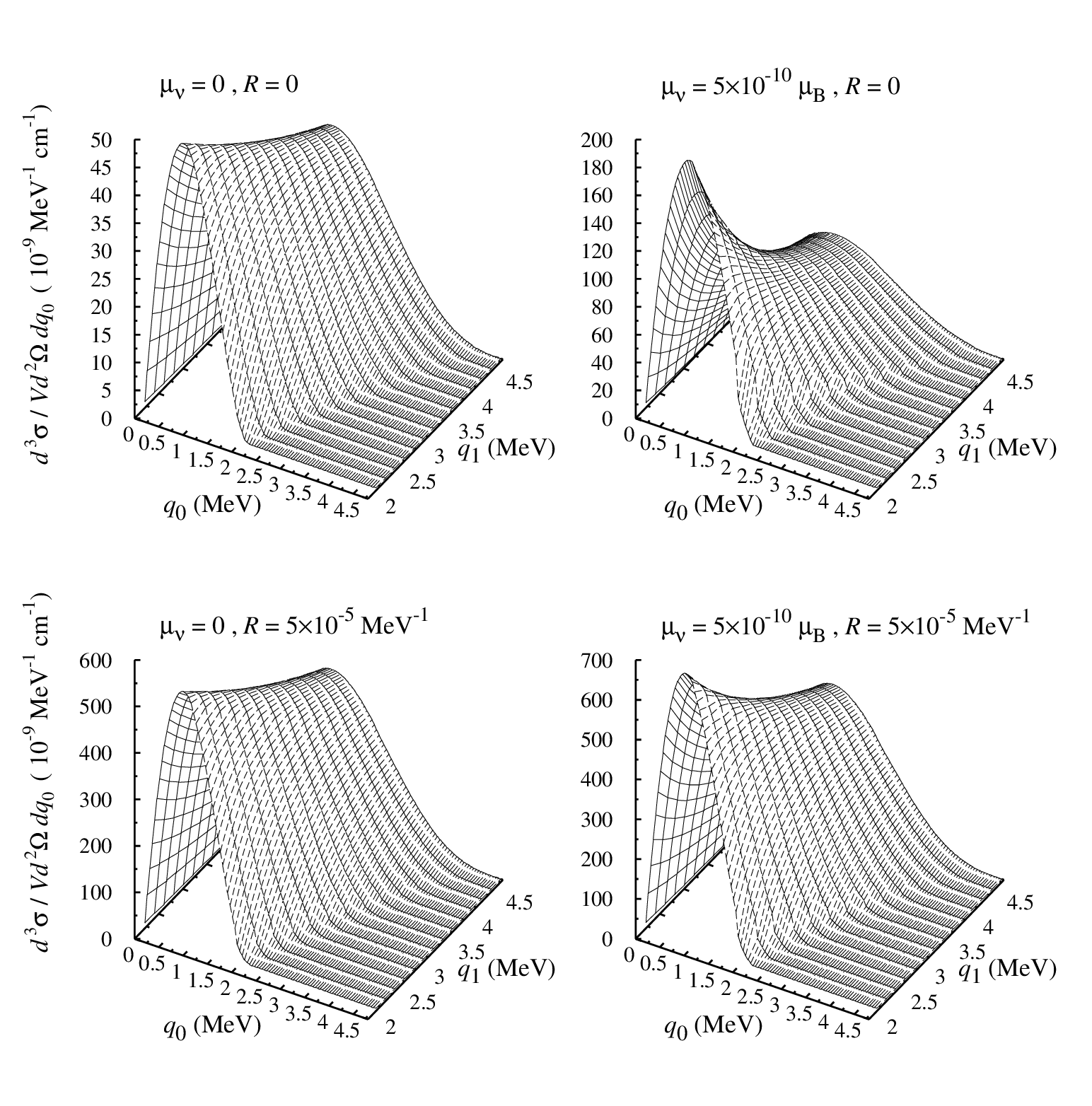,width=14.0cm}}
 \caption{ Same as in Fig.~\ref{ccpe5rb25}, but for  fixed baryon density ($\rho_B=5\rho_0$).}
\label{ccpe5rb5}
\end{figure*}

The left panel of Fig.~\ref{trap} shows the relative fraction of every constituent in the case that neutrinos are absent, while the right panel shows the case when neutrinos are trapped. We obtain a similar conclusion as in Ref.~\cite{chiapparini}, i.e., when Fermi momentum of the electrons reaches the muon mass, muon begins to appear, and then the proton and electron fraction curves are splitted into two different paths. The threshold for muon production occurs just below the saturation density. In the case that neutrinos are trapped, for example with $Y_{l_e}=0.3$, it can be seen that the  threshold is shifted toward higher densities. This displacement is even larger for the E-RMF model ($\rho=3 \rho_0$) than the result of Ref.~\cite{chiapparini} ($\rho=2 \rho_0$). This similar finding leads to a similar conclusion drawn in Ref.~\cite{chiapparini}, i.e., the EOS of matter with neutrino trapping is softer than the case where neutrinos are absent. This fact leads to a very important consequence for the physics of supernovae explosions~\cite{chiapparini}. 

Figures ~\ref{crossna} and ~\ref{crossnt} reveal the fact~\cite{caroline05} that a significant difference between total (including neutrino-electron electromagnetic properties) and weak differential cross sections  in electrons gas, starting more or less from $\mu_{\nu}> 10^{-10} \mu_B$ and $R>10^{-5}~{\rm MeV}^{-1}$  also occurs in dense matter. If neutrinos are present in matter, then the threshold values of  $\mu_{\nu}$ and  $R$ become more or less similar to the case where neutrinos are absent. The increment in the cross section right after the threshold is relatively faster for the neutrino trapping case compared to the  case where neutrinos are absent. The rapid increment of the transversal and longitudinal cross sections right after threshold seems to be the reason of this fact.

In Figs.~\ref{totccc} and ~\ref{totc1}, we show the total differential cross sections along with individual contributions of their constituents as a function of the energy transfer $q_0$ at a fixed momentum transfer $q_1=2.5$ MeV, neutrino energy $E_{\nu}= 5$ MeV and baryon density $\rho_B= 2.5 \rho_0$. 
In Fig.~\ref{totccc}, we set $R$ and $\mu_{\nu}$ to zero in left panels, while $\mu_{\nu}$= 5 $\times 10^{-10} \mu_B$ in the right panels.  The result in lower panels are obtained if neutrinos are absent, while those in upper panels are  obtained in the case where neutrinos are trapped. 
In Fig.~\ref{totc1} we set $\mu_{\nu}$= 0 and $R$= 5 $\times 10^{-5} {~\rm MeV}^{-1}$.  In  this figure the left panel exhibits the case where neutrinos are absent and the right panel shows the case where neutrinos are trapped.  

From the lower left panel of Fig.~\ref{totccc} (the standard weak interaction cases, where neutrinos are absent), we can see that the proton contribution is larger than the electron one but, on the contrary,  from the upper left panel of Fig.~\ref{totccc} (where neutrinos are trapped), the proton contribution is smaller than the electron one. Both contributions increase if the neutrinos are present, as shown in the upper left panel. It can also be seen that muons play almost no role in this case. However, since neutron contributions are dominant in both cases and they have more or less a same cross section magnitude, the difference  between neutrino absent and neutrino trapped in matter   in each individual contribution does not significantly show up in the total differential cross section.  

In order to see the effect more clearly, we take $\mu_{\nu}= 5 \times 10^{-10} \mu_B$ in the case of nonzero neutrino dipole moment. Obviously, this value is larger than its laboratory bound ($\mu_{\nu}=1.0 \times 10^{-10} \mu_B$) as well as  the bound from astrophysical consideration  ($3.0 \times 10^{-12} \mu_B$). The result can be seen in the lower right panel of Fig.~\ref{totccc} (for the neutrino absent case), where we can see that the proton contribution is larger than the neutron one and the electron contribution has a similar order to the neutron one, while the muons start to have a significant contribution. On the other hand, in the  neutrino trapping case (upper left panel of that figure), muons have almost no contribution. Contributions from protons and electrons are larger than those of neutrons in this case. The different number of particle distributions  of each constituent between both cases leads to a difference in the total differential cross sections. It  can also be seen in Fig.~\ref{totc1} that a similar situation also happens for the case of $R= 5.0 \times 10^{-5}~{\rm MeV}^{-1}$ but with  $\mu_{\nu}=0$.  

This means that in contrast to the calculation based on the standard weak interaction, the cross section calculated by including neutrino electromagnetic properties is very sensitive to the particle number of each constituent. 

To see how sensitive the calculated total cross section to the neutrino electromagnetic properties is, we plot the total differential cross section as functions of the energy transfer $q_0$ and momentum transfer $q_1$ for baryon density $\rho_B=2.5\rho_0$ in Figs.~\ref{cce5rb25} and ~\ref{ccpe5rb25}, and  for $\rho_B= 5 \rho_0$ in Figs.~\ref{cce5rb5} and ~\ref{ccpe5rb5}. Figures~\ref{cce5rb25} and ~\ref{cce5rb5} show the results in the neutrino absent  case, while  Figs.~\ref{ccpe5rb25} and ~\ref{ccpe5rb5} show the results in the case that neutrinos are present. For each figure, in the upper left panel we show the cross sections for fixed $\mu_{\nu}= 0$ and $R= 0$, in the upper right panel we set $\mu_{\nu}= 5.0 \times 10^{-10} \mu_B$ and $R= 0$, in the lower right panel $\mu_{\nu}=0$ and  $R= 5.0 \times 10^{-5}~{\rm MeV}^{-1}$, while in the lower right panel $\mu_{\nu}=5.0\times 10^{-10} \mu_B$ and $R= 5.0\times 10^{-5} { \rm MeV}^{-1}$.

Clearly from Fig.~\ref{cce5rb25}, for fixed  $\mu_{\nu}$, in general, the trends of cross sections between  $R=0$  and  $R=5.0\times 10^{-5}~{\rm MeV}^{-1}$ are quite similar. Only in the region of  $q_0\simeq 1-2$ MeV and   $q_1\simeq 2-4.5$ MeV, the shapes of cross sections seem to be quite different. The magnitudes of both cross sections are different due to the quite large value of $R$. However, in the case of fixed  $R$ and $\mu_{\nu}$ = 5.0  $\times 10^{-10} \mu_B$, the trend and magnitude of the total cross section change. It is found that the cross section decreases when  $q_1$ increases. If we see the  upper right panel of Fig.~\ref{cce5rb25}, it can be seen that this decrement can be slowed down if the charge radius is set to a non zero value (e.g., $5.0\times 10^{-5}~{\rm MeV}^{-1}$).  The effect of nonzero neutrino dipole moment appears more dominantly at smaller values of momentum transfer $q_1$ and energy transfer $q_0$, which is due to the role of massless photon propagators in the electromagnetic interaction. On the contrary, the weak contribution becomes more dominant at larger  values of $q_1$ and $q_0$. 

By comparing Fig.~\ref{cce5rb25} and Fig.~\ref{cce5rb5}, we can see that in higher densities  (i.e., $\rho_B=5.0 \rho_0$) the magnitude of each cross section becomes significantly large and, as a consequence, the difference between the total and the weak cross sections becomes more pronounced. Furthermore, the shapes of the cross sections become smoother in this case.

The case of  trapped neutrino (e.g., $Y_{l_e}=0.3$) is shown in Figs.~\ref{ccpe5rb25} and ~\ref{ccpe5rb5}, where we can see that the difference  between total and weak cross sections appears significantly due to the larger cross section, as we expected. Nevertheless, there is no indication that  the change in the cross section trend is due to the more pronounced difference of the role of each constituent in the higher density. The difference in the shapes of cross sections calculated by including and excluding electromagnetic form factors also appears in the region  of $q_0\simeq 1-2$ MeV and   $q_1\simeq 2-4.5$ MeV for low densities  (i.e., $\rho_B= 2.5\rho_0$), albeit with a different trend.  

In addition, if we considered the upper bound of the  neutrino-muon dipole moment which is given by $\mu_{\nu}$ $<$ 7.4  $\times 10^{-10} \mu_B$ ~\cite{allen,kraka}, besides the weak magnetism term as investigated by Ref.~\cite{Horo2}, it seems from the above discussion that the neutrino muon electromagnetic properties might give additional effects to the neutrino muon and its anti-neutrino mean free path difference in the neutron rich matter. How significant the effects should be quantitatively  checked by a real calculation. 

\section{Conclusions and Outlook}
\label{sec_sum}
In conclusion, we have studied the sensitivity of the neutrino cross section to the neutrino electromagnetic properties. In our calculations, we use the G2* parameter set of the E-RMF model to describe matter. This parameter set predicts a soft EOS at high density and has $\rho_B= 2.5\rho_0$ that coincide with the direct URCA process threshold. The calculation has been performed for the cases in which neutrinos are trapped and absent in matter. It is found that in the non strange stellar matter, the electromagnetic form factor has an important role in the neutrino-electron matter cross section if  $\mu_{\nu}>10^{-10} \mu_B$ and $R>10^{-5}~{\rm MeV}^{-1}$. It is  also found that the effects of the neutrino electromagnetic form factors on the cross sections are more pronounced at higher densities. 

Matters with trapped neutrino, like supernovae ones, are more sensitive  to the presence of neutrino electromagnetic properties. This is due to the fact that matters with trapped neutrino have larger fractions of protons and electrons than those without neutrinos.Although we have found that the role of neutrino electromagnetic properties in the neutrino-electron matter interaction is not too crucial, this would not be the case if we considered  the bound of the  neutrino-muon dipole moment given by Refs.~\cite{allen,kraka}.  

With increasing the density of the protoneutron star, strangeness can be liberated and face up in the the filling of the hyperon Fermi seas and/or in the creation of kaon condensates. Occurrence of these exotics in protoneutron star interiors will enhance the neutrino scattering rate and it may have also interesting observational consequences, like softening the EOS, possibility of changing nucleon isospin composition in the star matter evolution, as well as enhancing neutrino emission processes in the neutron star matter evolution, as extensively discussed in Refs.~\cite{kolo,knorren,baldo1,vidana1,bunta,zuo,banik,hua03,hua2}.  Further, depending on the temperature and the model used, baryon correlations can also reduce the scattering rate~\cite{Horo2,mornas2,horowitz91,caiwan,cowel}. Therefore, an extension of this calculation by including the strange matter, RPA correlation and a more general condition, i.e., finite temperature might also be interesting and quite relevant to consider in the future.
  
\appendix
\section{Lagrangian Densities}
\label{sec_app}

The  effective Lagrangian density of E-RMF model is ~\cite{Furnstahl96, Wang00}
\be
 {\mathcal L}^{\rm nuc}= {\mathcal L}_N + {\mathcal L}_M , 
\ee
where the nucleon part, up to order $\nu$ = 3, has the form of
\bea
{\mathcal L}_N &=&\bar{\psi}[i \gamma^{\mu}(\partial_{\mu}+i \bar{\nu}_{\mu}+i g_{\rho}  \bar{b}_{\mu}+i g_{\omega} V_{\mu})+g_A  \gamma^{\mu} \gamma^{5} \bar{a}_{\mu}\nonumber\\&-&M+g_{\sigma} \sigma]\psi-\frac{f_{\rho} g_{\rho}}{4 M }\bar{\psi}\bar{b}_{\mu \nu} \sigma^{\mu \nu}\psi,
\eea
with
\be
\psi=\left( {p \atop n}\right),  ~ ~ ~ ~\bar{\nu}_{\mu}=-\frac{i}{2}(\bar{\xi}^{\dagger}\partial_{\mu}\bar{\xi}+\bar{\xi}\partial_{\mu}\bar{\xi}^{\dagger})= \bar{\nu}_{\mu}^{\dagger},
\ee
\be
\bar{a}_{\mu}=-\frac{i}{2}(\bar{\xi}^{\dagger}\partial_{\mu}\bar{\xi}-\bar{\xi}\partial_{\mu}\bar{\xi}^{\dagger})= \bar{a}_{\mu}^{\dagger},
\ee
\be
\bar{\xi}= {\rm exp}(i \bar{\pi}(x)/f_{\pi}), ~ ~ ~ ~ \bar{\pi}(x)=\frac{1}{2} \vec{\tau}\cdot \vec{\pi}(x),
\ee
\be
\bar{\pi}(x)=\frac{1}{2} \vec{\tau}\cdot \vec{\pi}(x),
\ee
\be
\bar{b}_{\mu \nu} = D_{\mu}\bar{b}_{\nu}-D_{\nu}\bar{b}_{\mu}+i g_{\rho}[\bar{b}_{\mu},\bar{b}_{\nu}],  ~ ~ ~ ~  D_{\mu}=\partial_{\mu}+i\bar{\nu}_{\mu},
\ee
\be
V_{\mu \nu}=\partial_{\mu}V_{\nu}-\partial_{\nu}V_{\mu},
\ee
\be
\sigma^{\mu \nu}=\frac{1}{2}[\gamma^{\mu},\gamma^{\nu}].
\ee
Here, $p$, $n$  and $M$ are the proton field, neutron field and nucleon mass,  and  $\sigma$, $\vec{\pi}$, $V^{\mu}$, and $\vec{b}^{\mu}$ are the $\sigma$,  $\pi$, $\omega$ and $\rho$ meson fields, respectively.
The meson contribution, up to order $\nu$=4, reads
\bea
{\mathcal L}_M &=&\frac{1}{4}f_{\pi}^2 {\rm Tr} (\partial_{\mu}\bar{U}\partial^{\mu}\bar{U}^{\dagger})+\frac{1}{4}f_{\pi}^2 {\rm Tr}(\bar{U} \bar{U}^{\dagger}-2)+\frac{1}{2}\partial_{\mu}\sigma\partial^{\mu}\sigma\nonumber\\ &-&\frac{1}{2} {\rm Tr}(\bar{b}_{\mu \nu}\bar{b}^{\mu \nu})-\frac{1}{4}V_{\mu \nu}V^{\mu \nu}-g_{\rho \pi \pi}\frac{2 f_{\pi}^2}{m_{\rho}^2} {\rm Tr}(\bar{b}_{\mu \nu}\bar{\nu}^{\mu \nu})\nonumber\\
&+&\frac{1}{2}(1+\eta_1 \frac{g_{\sigma} \sigma}{M}+\frac{\eta_2}{2} \frac{g_{\sigma}^2 \sigma^2}{M^2})m_{\omega}^2 V_{\mu}V^{\mu}+\frac{1}{4!}\zeta_0 g_{\omega}^2 (V_{\mu}V^{\mu})^2\nonumber\\&+&(1+\eta_{\rho} \frac{g_{\sigma} \sigma}{M})m_{\rho}^2 {\rm Tr}(\bar{b}_{\mu}\bar{b}^{\mu})-m_{\sigma}^2\sigma^2(1+\frac{\kappa_3}{3 !} \frac{g_{\sigma} \sigma}{M}+\frac{\kappa_4}{4 !} \frac{g_{\sigma}^2 \sigma^2}{M^2}),
\eea
where
\be
\bar{U}=\bar{\xi}^2,  ~ ~ ~ ~\bar{\nu}_{\mu \nu} = \partial_{\mu}\bar{\nu}_{\nu}-\partial_{\nu}\bar{\nu}_{\mu}+i [\bar{\nu}_{\mu},\bar{\nu}_{\nu}]=-i[\bar{a}_{\mu},\bar{a}_{\nu}].
\ee
In the mean field approximation, $\pi$ meson does not have any contribution. 
If we set $\eta_1$,  $\eta_2$, $\zeta_0$, $\eta_{\rho}$ and $f_{\rho}$ equal to zero, we will obtain the same nucleon and meson  equations as in the standard RMF models~\cite{pg,ring,serot}.

To achieve a softer density dependence of the nuclear matter symmetry energy of  the standard RMF model, Refs.~\cite{Horowitz01,Shen05} add  isovector-vector nonlinear terms in the Lagrangian density. In this paper, a similar procedure as in Refs.~\cite{Horowitz01,Shen05} is adopted. Since in the  E-RMF model the isovector vector nonlinear term is already present, the  density dependence of the nuclear matter symmetry energy can be adjusted without adding new isovector nonlinear terms. Thus, we only adjust $g_{\rho}$ and  $\eta_{\rho}$ but maintain the requirement that the symmetry energy at $k_F =1.14$ fm should have the same value at $E_{\rm sym}=24.1$ MeV. The argument behind this procedure is  explained in detail in Refs.~\cite{Horowitz01,Shen05}.
For leptons, we use the following  free Lagrangian density:
\be
\sum_{l\,=\,e^-,\,\mu^-,\,\nu} \bar{l}( \gamma^{\mu}\partial_{\mu}-m_l)l.
\ee

\section*{ACKNOWLEDGMENT}
We acknowledge P. Danielewicz for his kindness to give us  his experimental data.

\end{document}